\documentclass{article}
\usepackage{arxiv}
\usepackage[utf8]{inputenc} 
\usepackage[T1]{fontenc}    
\usepackage{hyperref}       
\usepackage{url}            
\usepackage{booktabs}       
\usepackage{nicefrac}       
\usepackage{microtype}      
\usepackage{lipsum}

\usepackage{cite}
\usepackage{amsmath,amssymb,amsfonts}
\usepackage{pifont}
\usepackage{multirow}
\newcommand{\cmark}{\ding{51}}%
\newcommand{\xmark}{\ding{55}}%
\usepackage{algorithmic}
\usepackage{graphicx}
\usepackage{textcomp}
\def\BibTeX{{\rm B\kern-.05em{\sc i\kern-.025em b}\kern-.08em
    T\kern-.1667em\lower.7ex\hbox{E}\kern-.125emX}}
    
\title{A Consent Model for Blockchain-based Distributed Data Sharing Platforms}

\author{
  Vikas Jaiman \\
  Institute of Data Science (IDS) \\
  Maastricht University \\
   6211 LK Maastricht, The Netherlands \\
  \texttt{v.jaiman@maastrichtuniversity.nl} \\
   \And
 Visara Urovi \\
  Institute of Data Science (IDS)\\
  Maastricht University\\
  6211 LK Maastricht, The Netherlands \\
  \texttt{v.urovi@maastrichtuniversity.nl} \\
}

\begin{document}
\maketitle

\begin{abstract}
In modern healthcare systems, being able to share electronic health records is crucial for providing quality care and for enabling a larger spectrum of health services. Health data sharing is dependent on obtaining individual consent which, in turn, is hindered by a lack of resources.
To this extend, blockchain-based platforms facilitate data sharing by inherently creating a trusted distributed network of users. These users are enabled to share their data without depending on the time and resources of specific players (such as the health services). In blockchain-based platforms, data governance mechanisms become very important due to the need to specify and monitor data sharing and data use conditions. In this paper, we present a blockchain-based data sharing consent model for access control over individual health data. We use smart contracts to dynamically represent the individual consent over health data and to enable data requesters to search and access them. The dynamic consent model extends upon two ontologies: the Data Use Ontology (DUO) which models the individual consent of users and the Automatable Discovery and Access Matrix (ADA-M) which describes queries from data requesters. We deploy the model on Ethereum blockchain and evaluate different data sharing scenarios. The contribution of this paper is to create an individual consent model for health data sharing platforms. Such a model guarantees that individual consent is respected and that there is  accountability for all the participants in the data sharing platform. The evaluation of our solution indicates that such a data sharing model provides a flexible approach to decide how the data is used by data requesters. Our experimental evaluation shows that the proposed model is efficient and adapts to personalized access control policies in data sharing.
\end{abstract}

\keywords{Blockchain \and Data sharing \and Distributed ledgers \and EHR exchange \and Individual consent}

\section{Introduction}
\label{sec:introduction}
With the rapid development of big data technology, enormous amounts of data are currently being generated~\cite{datacreate}. Data providers such as companies, organizations, and individuals increasingly chose to share their data for research and other various purposes. As data sharing rapidly steps up, sharing practices are lagging in specifying and protecting individual consent preferences. 

\noindent \textbf{Consent Matters.} 
In this work, we enable individuals to decide how to share their health data with data requesters, such as researchers, organizations, or companies. Although distributed internet-based solutions (such as cloud-based systems) can already connect data providers and data requesters, the difficulty of ensuring compliance from both parties remains. Individual data sharing faces the problem of unambiguously defining consent while data requesters face the issue of clearly stating their intentions (i.e. A well-known example of this conflict is the Cambridge Analytica case~\cite{cambridge}). Furthermore, storing data in internet-based solutions (such as cloud-based systems) creates large pools of data that are handled almost exclusively by a few big corporate players with business incentives to maintain this status quo~\cite{van2019reframing}. This raises data privacy concerns, especially when the data consist of sensitive health records. For example, cloud-based systems store large data aggregates in data centers. 
In order to move towards decentralized data sharing models, we need to enable individual users to steer data sharing practices. In this paper, we propose a generic methodology to dynamically represent individual consent for data-sharing.

\noindent \textbf{Challenges.} 
The problem with the ownership and privacy of personal health data is exemplified by the issue of data generated by consumer wearables. Even if wearables collect individual data, users do not typically own these data. Instead, the data are collected and stored by device manufacturers ~\cite{piwek2016rise} who often charge a monthly fee for the use of their platform~\cite{venkataramanan2014my}. By contrast, in a patient-centered data sharing ecosystem, data is available to the user; it is controlled by the person who generated it; it is privacy-enabled and secure, and maintains data provenance information~\cite{kish2015unpatients}. Moreover, the  General Data Protection Regulation (GDPR)~\cite{gdpr} adopted by the European Union, suggests that default entitlements in personal data rest with the data subject and so do the individual control rights. Amongst these rights, GDPR is giving individuals the right to restrict processing and to request deletion of data which is known as the “right to be forgotten”~\cite{purtova2015illusion}. To this extend, blockchain is a novel technology that enables fine-grained data from each individual to create a global resource where data integrity and immutability of transaction is ensured. Blockchain enables the storage of transactions relating to data exchange in a verifiable and non-reputable way by all the actors in the system. At the core of blockchain, systems are the decentralization of control. Decentralization offers a substantial opportunity in a data-sharing scenario because it eliminates centralized control over the data and it enables individuals to decide how to share their personal records. 
Although platforms to support data sharing with blockchain-based solutions have been proposed~\cite{opf, provchain, dataprov, medrec} to date, there are no frameworks that model individual consent in relation to the purpose of use. In order  to construct a dynamic consent-based model we need to address the following questions: 
\begin{itemize}
    \item How can individual consent be represented in a machine-readable way?
    \item How the users will be connected over the platform?
    \item How we can dynamically match the purpose of the use of a requester query to the individual consent of data providers?
    \item How we can ensure data integrity and maintain a verifiable history of changes over time?
   \item How we can build an efficient framework? 
\end{itemize}   
\noindent \textbf{Contributions.} To address the above issues, we introduce a dynamic consent model for health data sharing using blockchain technology. Individuals, as data providers, decide on the data use conditions and access that data requesters must comply with.  Such a model guarantees that individual consent is respected and that there is accountability for all the participants in the data-sharing platform. It provides a GDPR compliant model, whereby all the uses of data are monitored by the decentralized model, and permissions can be modified or removed upon the users' request.
The novelty of the solution stands in combining two consent representation models (DUO~\cite{DykeDUO} and ADA-M~\cite{WoolleyADAM}) to represent and standardize user consent and to dynamically match it to data requester purpose statement queries. 
By using a blockchain model, we maintain a transaction history of all transactions related to data. Through this, the data provider specifies individual data sharing rules, monitors the use of data, and revokes or updates access to data anytime. 

The remainder of this paper is structured as follows: Section~\ref{sec:prelimiaries} discusses the background work related to our data-sharing platform.
Section~\ref{sec:relatedwork} discusses related work and highlights the limitations of state-of-the-art. Section~\ref{sec:architecture} presents the consent-based architecture. Section~\ref{sec:implementation} presents the implementation of the solution. Next, Section~\ref{sec:evaluation} presents the evaluation of the solution via LUCE~\cite{luce}. Section~\ref{sec:discussion} discusses the impact of such solution in terms of scalability, privacy and data heterogeneity. Finally, section~\ref{sec:conclusion} presents the conclusion and future work.

\section{Background}\label{sec:prelimiaries}
In this section, we briefly review the background work on data-sharing with blockchain technology, smart contracts, and consent codes used for consent representation.

\subsection{Data Exchange with Blockchain Technology}
\begin{figure}[t!]
\centering
\includegraphics[width=0.9\textwidth]{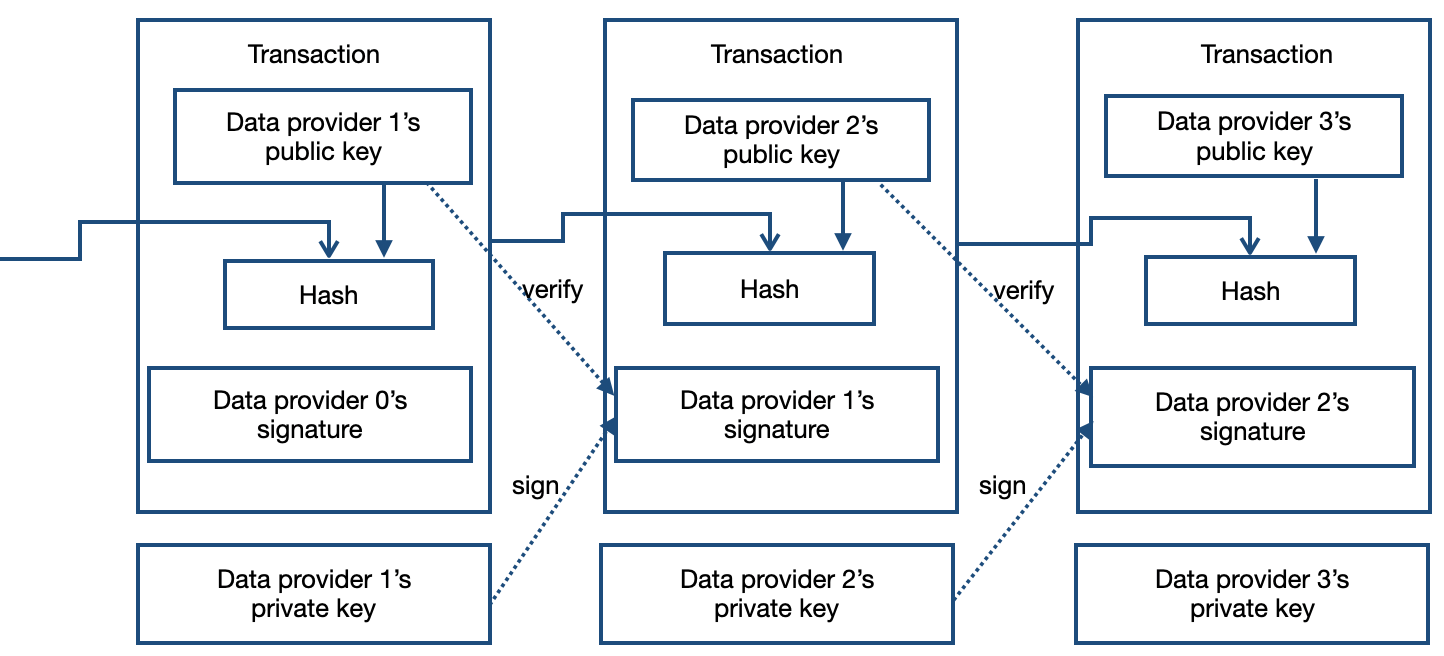}
\caption{Chain of transactions in blockchain.}
\label{fig:blockchain}
\end{figure}

Blockchain was initially proposed by Nakamoto \textit{et al.}~\cite{nakamoto2008bitcoin} in 2008 as a digital cryptocurrency which is based on a distributed ledger technology as seen in Fig.~\ref{fig:blockchain}. The model is permissionless, meaning that anyone can join and leave the network. The interaction occurs in the form of transactions that are bundled together in blocks. When a transaction is generated, it is broadcasted to the network where it propagates. Adding valid transactions and appending a new block in a blockchain is performed by specific nodes, called miners. In order to add a new block to the blockchain, miners have to solve a cryptographic puzzle. Depending on the blockchain model, solving the puzzle can be quite expensive in terms of computational cost. 
The first miner who solves the puzzle broadcasts the new candidate block to the network together with the solution of the puzzle. The consensus among nodes is reached by verifying that the identified solution is valid. Finally, the candidate block is added to the chain. As seen in Fig.~\ref{fig:blockchain}, the blocks in the chain are linked to their predecessor block through a unique hash value. Through this, all the network shares the same transaction data with the certainty that, the information stored in the chain contains valid transactions that are verified and cannot be tampered.
Nowadays, blockchain technology has evolved from a crypto-currency solution into more general-purpose blockchain platforms. Ethereum~\cite{ethereum} is one of the widely used general-purpose blockchain solutions. It uses an Ethereum virtual machine (EVM) to run programs called \textit{smart contracts}. Smart contracts incorporate the business logic of transactions, thereby defining when and what constitutes a valid transaction.

Blockchain enables decentralized data sharing with no need for centralizing the data to a particular organization. In this way information on data transactions between data providers and data requesters are immutable and safeguard the data ownership. Via blockchain, it is possible to monitor the data exchange and keep a data transaction history across the nodes in a distributed leaderless manner.
\subsection{From Data Sharing to Smart Contracts}
 Smart contracts~\cite{smartcontract,SmartcontractoverviewZheng20} are self-executable codes that control the data transfer or cryptocurrencies over the blockchain network. Smart contracts translate contractual clauses into a code that enables automated workflows and eliminate the need for a trusted third party. 
  
 A smart contract executes independently and automatically based on the data that was entered in the triggering transaction. The most prominent blockchain that supports smart contracts is the Ethereum blockchain. In this paper, we use Ethereum smart contracts~\cite{smartcontractethereum} to model the interaction between individuals, which we refer to as data providers, and data requesters. The two actors interact with each other via smart contracts. These are configured by data providers to capture the logic for dynamically deciding if the data can be shared with data requesters.
   
\begin{figure}[t!]
\centering
\includegraphics[width=0.9\textwidth]{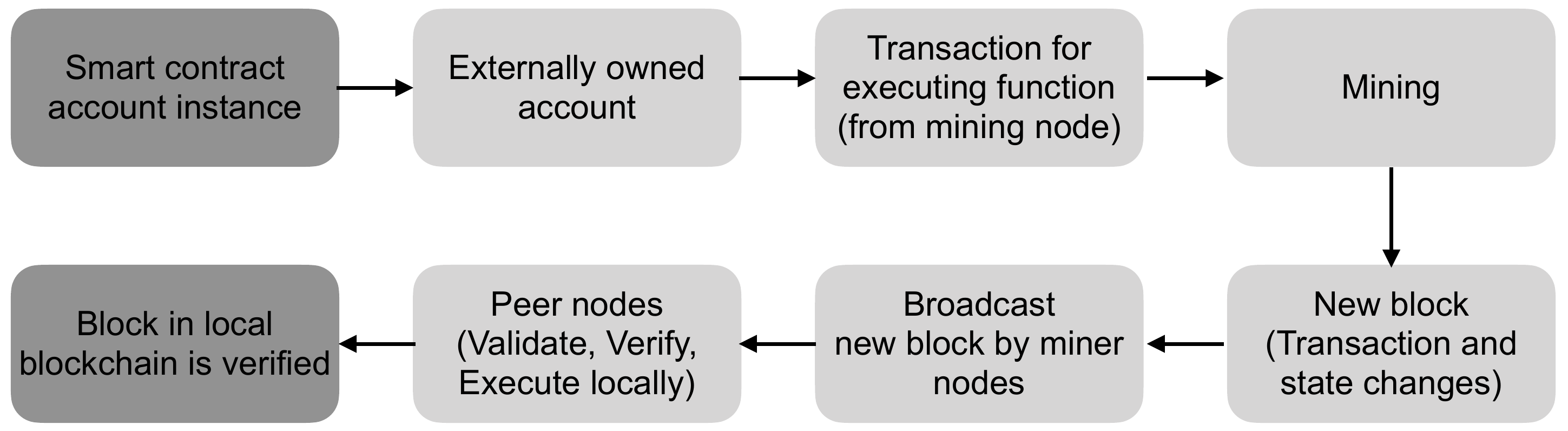}
\caption{Smart contract execution in Ethereum.}
\label{fig:contractexecution}
\end{figure}

Fig.~\ref{fig:contractexecution} represents the execution of a smart contract in Ethereum. A smart contract contains functions that can be executed by \textit{external account}. To execute a function, the decentralized application retrieves an instance of a smart contract by its address and starts a transaction. To take effect, a transaction has to be mined by peer nodes. After successful mining, a new block containing the transaction address is created. The miner node broadcasts the new block to the peer nodes. Afterwards, this new block is validated, verified, and executed by peer nodes and added to the blockchain.
\subsection{Data sharing with LUCE}
Havelange \textit{et al.}~\cite{luce} developed LUCE, a blockchain-based data-sharing platform to automatically track compliance with data licensing terms, and to facilitate data sharing concerning the rights of data subjects. To resolve trust issues between the two data sharing parties, LUCE attaches a license to the data. The license clearly states data usage terms. Via the blockchain platform, LUCE automatically tracks compliance with the licensing terms, even after the data has been acquired by the data requester. The use of data from data requesters is recorded by LUCE, and if a data requester acts in a way that doesn't comply with the data requester terms, the data provider can act on his rights. For example, data subjects can enforce GDPR rights to access, rectify, and erase the data. LUCE distinguishes between three different roles: Data Providers, Data Requesters, and Supervisory Authorities~\cite{luce}.
By building on a blockchain solution LUCE offers the following features:
\begin{itemize}
    \item Immutable information: Transactions relating to data exchange are stored in the platform and can not be destroyed. Therefore, the transactions showing which data provider shared data and which data requester has access to specific data remain available indefinitely in the platform.
    \item  Ownership: The smart contract specifies data access and links to the owner, therefore, a data provider can prove the existence and integrity of data processed on the blockchain.
    \item Transparency: A shared distributed ledger of all data usage is created. Knowing that the ledger can be changed only by having consensus, the transactions are known to be consistent, transparent, and accurate. The supervisory authority can inspect the transactions and intervene accordingly.
\end{itemize}
In this paper, the data provider uploads individually owned data to LUCE and specifies the terms of use. We extend LUCE with a consent model and test our model with data generated through IoT devices. \textit{Data requesters} use LUCE to find datasets based on their needs. The supervisory authority is responsible for intervening on data misuse and for enforcing GDPR rights. In this paper, we introduce the consent representation model and test the model with IoT data that are shared in the LUCE platform.
\subsection{Consent codes in Data Sharing}\label{sec:consentmanagement}
Ontologies provide a conceptualization of domain knowledge such that the information can be understood and processed by machines and by humans. Ontologies facilitate the interpretation of consent in regards to data use conditions. Two ontologies were considered relevant for modeling individual consent for supporting EHR data sharing:
\subsubsection{Data Use Ontology (DUO)}
Dyke \textit{et al.}~\cite{DykeDUO} provides an ontological representation of data use categories using \textit{consent codes}. It includes terms describing data use conditions for research data in the health/clinical/biomedical domain. 
\begin{table}
\caption{The primary categories of DUO. Data providers use them to express data-sharing consent preferences~\cite{DykeDUO}.}
\label{table:consentstatement}
\setlength{\tabcolsep}{3pt}
\def\arraystretch{1.5}
\begin{tabular}{p{230pt}p{200pt}}
\hline
 \textbf{Main Categories} & \textbf{Meaning/Assumptions} \\
\hline
\textbf{Primary Category (Mandatory)} &   \\
\hline
No restrictions (NRES) & $\bullet$ No restrictions on using the data  \\
\hline
\multirow{2}{*}{\begin{minipage}{3in}General research use and clinical care (GRU-CC)\end{minipage}}& $\bullet$ For health, medical or biomedical purposes \\
    & $\bullet$ Includes population origin or ancestry study\\
\hline
\multirow{2}{*}{\begin{minipage}{3in}Health, Medical or Biomedical research and clinical care (HMB-CC)\end{minipage}}& $\bullet$ Limited to HMB purposes \\
    & $\bullet$ Excludes population origin or ancestry study\\
\hline
\multirow{2}{*}{\begin{minipage}{3in}Open the dataset to population and ancestry research (POA)\end{minipage}}& $\bullet$ Use of the data limited to study of population origins or ancestry \\
\hline
Disease specific research and clinical care (DS-[XX]-CC) & $\bullet$ Use of the data must be related to a disease \\
\hline
\end{tabular}
\end{table}
Specifically, DUO standardizes data use restrictions into machine-readable consent.  Being a simple model, it is easy for data providers to define their consent. DUO, however, from the data requester perspective, lacks the details needed to represent the purpose of use. This is why we use the DUO model only to define individual consent choices of data providers. 

\noindent\textbf{Consent Statement.}
Table~\ref{table:consentstatement} shows an abstract view of the DUO model. Initially, a dataset is assigned to a primary category according to the data provider's consent. More restrictions on individual data can be included by specifying more categories. These are classified as secondary categories and can only be selected after choosing one of the primary categories. Examples of such restrictions include Non-profit-only, geographical restrictions, time limits, etc.
\subsubsection{Automatable Discovery and Access Matric (ADA-M)}
Woolley \textit{et al.}~\cite{WoolleyADAM} define a matrix of categories for expressing data use restrictions. This is a comprehensive information model that provides the basis for producing structured metadata \textit{profiles}. ADA-M is a complementary approach for classifying data use conditions and permissions.
In our model, we take advantage of the granularity of the ADA-M model for representing the purpose statements of data providers.

\noindent\textbf{Purpose Statement.} As shown in Table~\ref{table:purposestatement}, the ADA-M model~\cite{WoolleyADAM} uses categories e.g. Research Purpose and sub-categories e.g. use for biomedical research (HMB) as a way to represent consent with more detail.
\begin{table}
\caption{ADA-M~\cite{WoolleyADAM} categories for representing purpose statements of data requesters.}
\label{table:purposestatement}
\setlength{\tabcolsep}{3pt}
\def\arraystretch{1.5}
\begin{tabular}{p{230pt}p{200pt}}
\hline
 \textbf{Main Categories} & \textbf{Subcategories} \\
\hline
\multirow{2}{*}{\begin{minipage}{3in}Research Purpose (GRU)\end{minipage}}& $\bullet$ Use for methods development (NMDS) \\
    & $\bullet$ Use for reference or control material (RS-XX)\\ & $\bullet$ Use for research concerning populations (PO) \\ & $\bullet$ Use for research ancestry (ANS) \\ & $\bullet$ Use for biomedical research (HMB) \\
\hline
 \multirow{2}{*}{\begin{minipage}{3in}Health, Medical or Biomedical research Purpose (HMB)\end{minipage}}& $\bullet$ Use for research concerning fundamental biology (FB) \\
    & $\bullet$ Use for research concerning genetics (GSO)\\ & $\bullet$ Use for research concerning drug development (DD) \\ & $\bullet$ Use for research concerning any disease (DS-XX) \\ & $\bullet$ Use for research concerning age categories (AGE) \\ & $\bullet$ Use for research concerning gender categories (GEN) \\
\hline
\multirow{2}{*}{\begin{minipage}{3in}Clinical Purpose (CC)\end{minipage}} & $\bullet$ Use for decision support (DSO) \\ & $\bullet$ Use disease support (DS) \\
\hline
\end{tabular}
\end{table}
\section{Related Work} \label{sec:relatedwork}
Several research works~\cite{biobankconsent}~\cite{consentRantos} on consent based data sharing have been proposed in different domains. Riggs \textit{et al.}~\cite{consentgenomicdata} address the issue of data sharing with consent in genomic data. They introduce a one page web-based consent form for genomic data sharing which avoids multi-pages consent terms. Via a questionnaire with 5162 respondents, they show improved data access when the simplified consent form is used. 
In the privacy domain, authors~\cite{consentmaler} 
propose a web standard protocol called User-Managed Access (UMA) to enable applications to offer stronger consent management abilities. 
UMA gives an individual a unified control point for authorizing who and what can get access to his or her online personal data and services (such as creating status updates), no matter where those resources live online. While a generic solution and versatile for online data, UMA does not explicitly solve consent requirements from an EHR data-sharing perspective as well it suffers from lack of audibility over data access. Indeed the intentions of the data requesters, when they gain data access, are not explicit.
Through our solution, we categorize the consent according to the terms of use. The audibility of the platform enables the identification of any inappropriate data use, the revocation of access and the possibility to be forgotten by the data requester at any time. 
Several blockchain-based solutions~\cite{datasharingthwin}\cite{datasharingblockchain}\cite{datasharingzheng}  have been recently proposed for data sharing. These solutions focus on privacy-preserving data sharing, data management of EHRs, or sharing of health data on cloud storage. While these approaches can potentially improve data sharing practices, so far only a few address to some extent how consent could be included in data-sharing blockchain platforms.
In particular, Bhaskaran \textit{et al.}~\cite{bhaskaran} propose a consent based double-blind anonymous data sharing built on Hyperledger Fabric~\cite{hyperledger}. Their model focuses on defining a  Know your Customer (KYC) application which is built on a permissioned blockchain. Differently from our solution, this model is not suitable in a permissionless setting where any user can join and contribute to the platform. Moreover, in contrast with our model which enables both data providers and data requesters to dynamically define the terms of data usage, the purpose of the use is not modeled for data requester queries. Thus, the proposed model of Bhaskaran \textit{et al.}~\cite{bhaskaran} is currently unable to dynamically adapt the consent to different data queries. 
ProvChain~\cite{provchain} and DataProv~\cite{dataprov} are blockchain-based approaches to verify data provenance accountability. 
These solutions however are not focused on modeling dynamic consent. Neisse \textit{et al.}~\cite{neisse} propose a blockchain-based approach for data accountability and data provenance tracking. Their solution enables data tracking and reuses according to the data provider's consent. The solution includes a smart contract that describes the terms of use as well as data provenance information. In contrast to our model, this solution mainly focuses on data providers and data controllers but does not meaningfully include data processors such as researchers who are reusing the data by maintaining the individual consent rights of the data subjects. 
\begin{figure}[t!]
\centering
\includegraphics[width=0.95\textwidth]{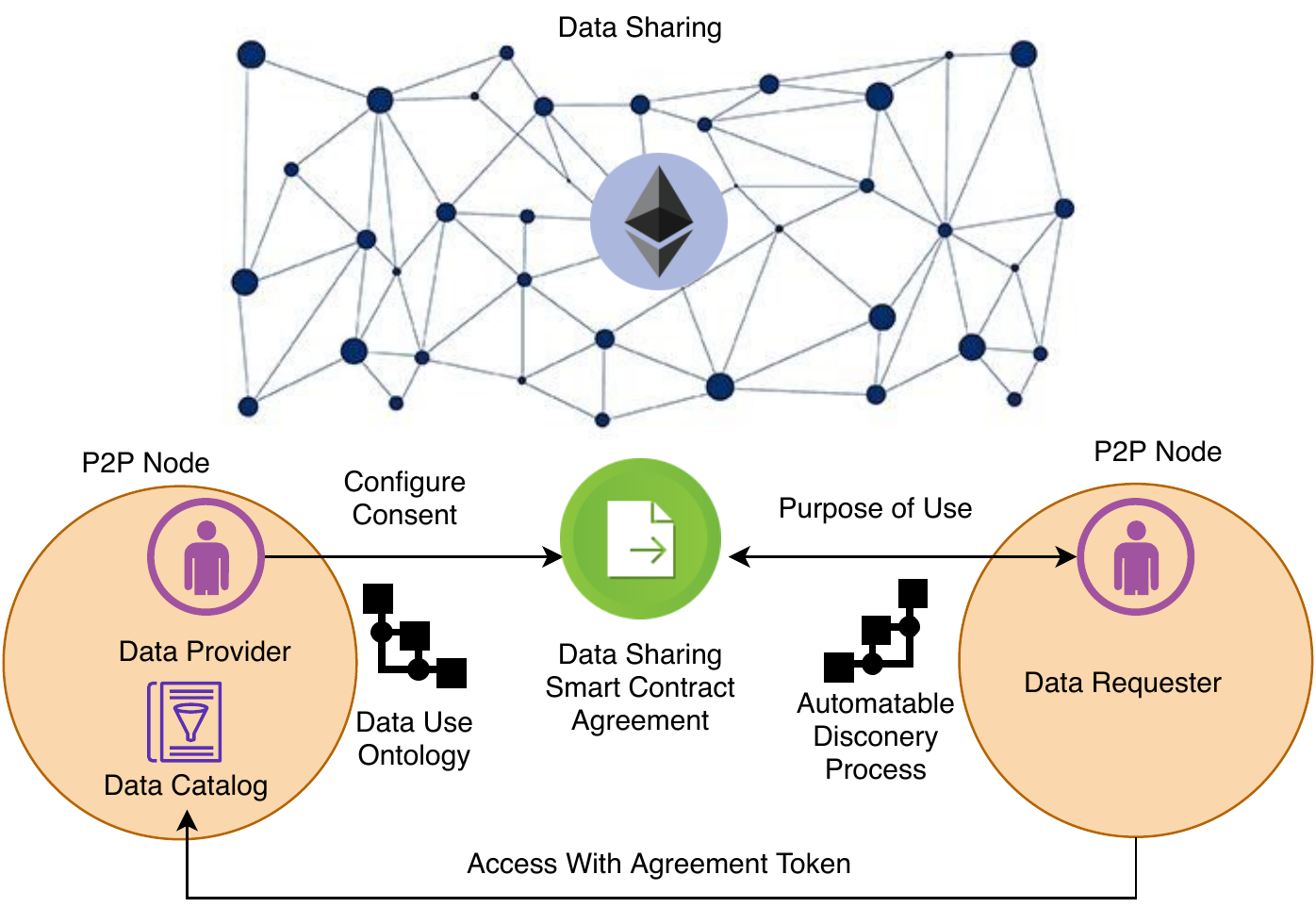}
\caption{A consent model for blockchain based architecture using smart contracts.}
\label{fig:architecture}
\end{figure}
\begin{figure}[t!]
\centering
\includegraphics[width=0.9\textwidth]{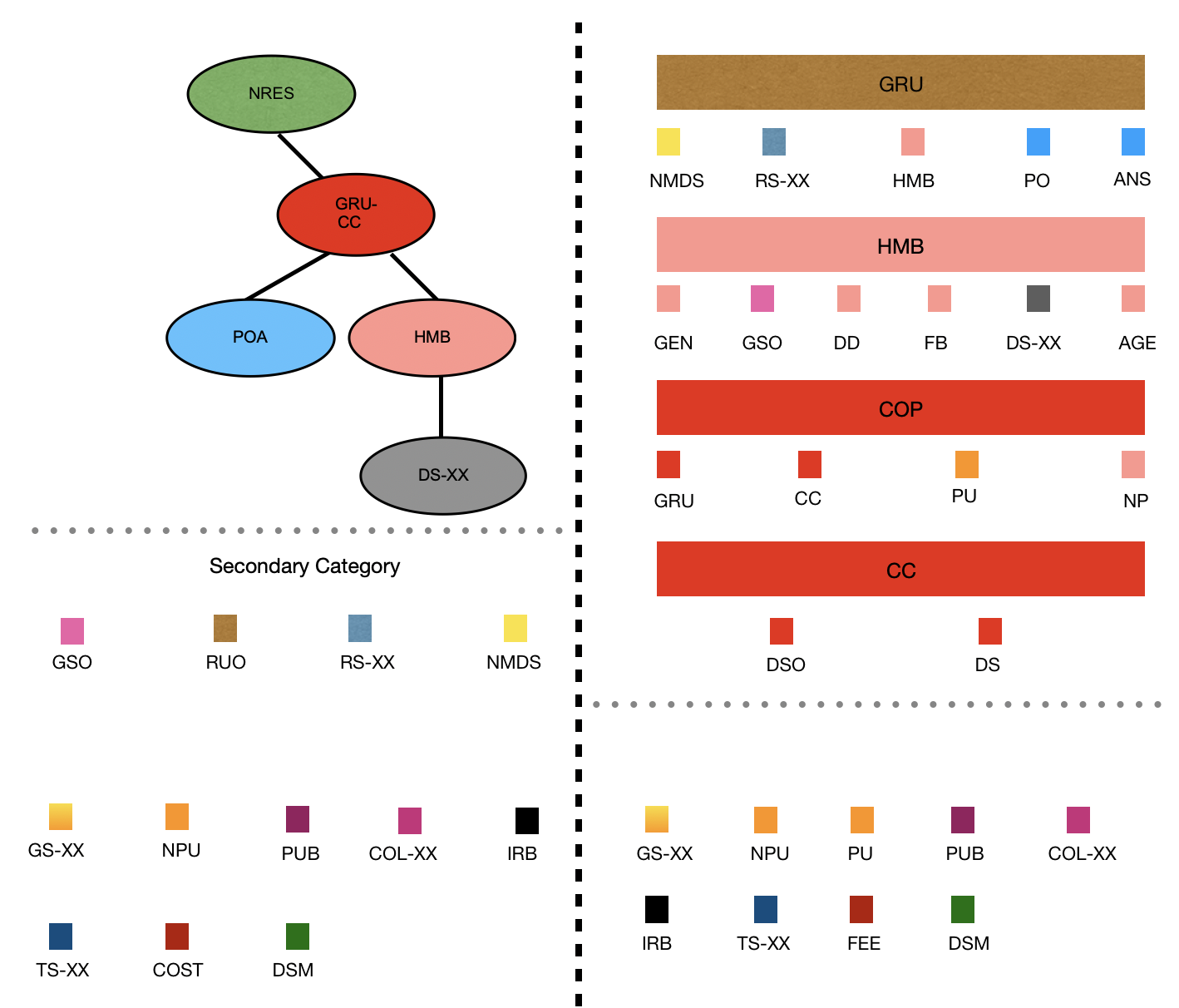}
\caption{Consent codes matching for DUO (left-side) and ADA-M matrix (right- side).}
\label{fig:consentmatchng}
\end{figure}
\section{Consent based Architecture Model}\label{sec:architecture}
In this section, we illustrate the architecture of our framework.
Following and extending upon the LUCE architecture~\cite{luce}, we distinguish the following actors and resources. 
\begin{itemize}
    \item Data provider: A data provider owns and shares the data. A data provider can specify consent in plain text form by using DUO terms. Data providers can be any individual, researcher, or an organization with data ownership. The key functionality of the data provider is: a) To register in the LUCE platform b) Publish data with the desired consent c) Update data if required d) Remove data entirely if desired. 
    \item Data requester: A data requester is looking for data for a specific purpose. A data requester can request the data from a data provider by using the ADA-M matrix form. The key functionality of data requester is:
    a) To register on the LUCE platform b) Search for the data by giving a purpose statement c) Request access to the data once find the desired one d) Confirm compliance with consent terms of the data provider.
    \item EHR data: EHR data (Electronic health records) contains information about the medical history of a patient, it is owned by a patient, authorized practitioner or an organization\footnote{We extend the EHR data definition to data collected from wearables as they contain valuable information on the physiological parameters of patients. }. 
    \item Datastore node: A datastore node stores all the information related to the required data. It could be cloud storage or local organizational storage.\\
\end{itemize}

Additionally, we work under the following assumptions: 
\begin{itemize}
    \item Datasets are shared as a whole. Thus, in the current model, a data provider is unable to share only a portion of a dataset.
    \item Data records are anonymized by data providers prior to enabling data-sharing.
\end{itemize}

The workflow of a consent model using blockchain is depicted in Fig.~\ref{fig:architecture}. The data provider gives the consent using the DUO model as specified in Fig.~\ref{table:consentstatement}. The consent statement of the data provider is recorded into the blockchain using a smart contract. When a data requester queries the data, it uses the ADA-M matrix ontology. The \textit{purpose statements} of data requesters are recorded into the blockchain via the smart contracts. This occurs when a smart contract matches the consent of a data provider (\textit{consent statement}) to a query from the (\textit{purpose statement}) of a data requester. 
   
Fig.~\ref{fig:consentmatchng} represents the matching of consent codes between the DUO and ADA-M matrix. It represents the DUO consent codes with abbreviations that match the ADA-M codes. In the figure, a similar color signifies that there is a consent match between two parties. For example, as specified in Table~\ref{table:consentstatement}, if a data provider gives consent for HMB research (HMB), this is represented in solid pink color in Fig.~\ref{fig:consentmatchng} (left-side). Similarly, if there is a query regarding Fundamental biology (FB) research we can observe in Fig.~\ref{fig:consentmatchng} (right-side) that \textit{FB} has a solid pink color and there is a match in the given consent. If there is a match, access is granted to the data and the datastore is accessible to the data requester. In case of no match, the request is denied and the data requester is notified.
\section{Consent Model Implementation}\label{sec:implementation}
In this section, we provide the implementation details of the smart contract and the extension of LUCE~\cite{luce} with the consent model proposed in this paper. 

\noindent\textbf{Experimental setup.} We implement the smart contract in Solidity~\cite{solidity}, a language for smart contracts provided by the Ethereum blockchain. Our consent model is then deployed into the LUCE~\cite{luce} platform --- a blockchain-based data sharing platform implemented in Python and deployed on the Ganache Ethereum network~\cite{ganache}. To run our experiments, we use a LuceVM virtual machine~\cite{luce} which is running on a 64 bit Ubuntu 16.04 LTS (Xenial Xerus) Linux operating system. The virtual machine is equipped with 1024 MB RAM. Our consent model implementation is available as open-source\footnote{https://gitlab.com/vjaiman/consentblockchainluce}. 
 
LUCE is implemented on top of the Ethereum blockchain. It uses Web3 javascript libraries~\cite{web3js} to interact with the Ethereum blockchain. LUCE uses Django~\cite{django} for implementing the user interface. The data providers interact via the Django web framework to share the data and to specify the associated consent. LUCE stores the link between the smart contract and the corresponding datastore location.
 
Every data provider has a smart contract that is used to record the consent and to compare it to the data requester's purpose of use statement. 
\begin{figure}[t!]
\centering
\includegraphics[width=0.95\textwidth]{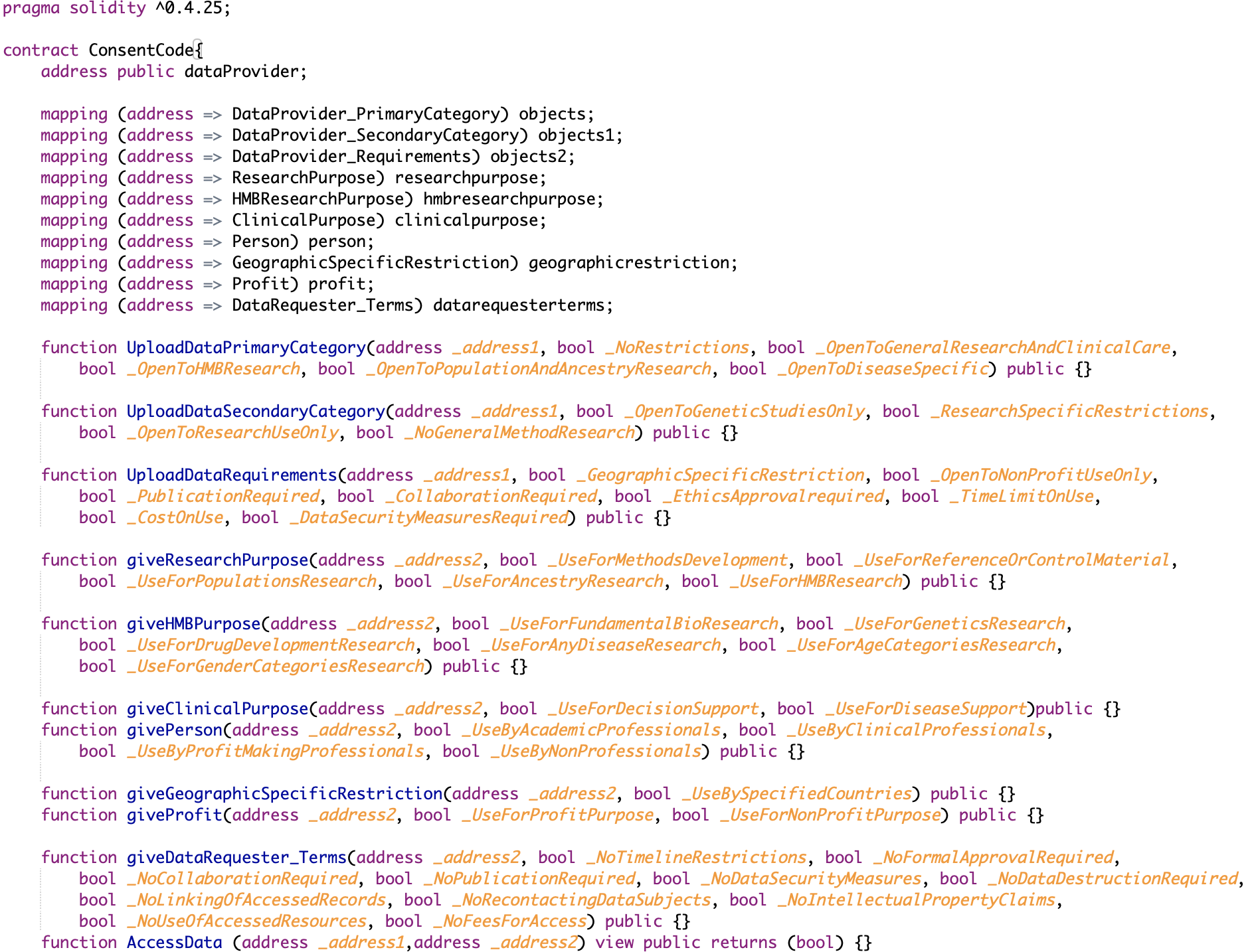}
\caption{Smart contract in consent model.}
\label{fig:codesnippet}
\end{figure}
Fig.~\ref{fig:codesnippet} shows the code snippet of a smart contract for our data sharing platform. The contract stores the  data provider consent statement as a \textit{struct} named \textit{DataProvider\_PrimaryCategory}. This describes the consent given by the data provider while publishing the data. Optionally, data providers can specify \textit{DataProvider\_SecondaryCategory} and \textit{DataProvider\_Requirements}, if more restrictions are needed on the data consent. By calling the \textit{UploadDataPrimaryCategory} function, the consent of the data provider is stored on the blockchain. Similarly, data provider can call the \textit{UploadDataSecondaryCategory} and \textit{UploadDataRequirements} functions to submit the additional terms for data use. Data requester function \textit{giveHMBpurpose} requests the data by submitting the purpose of use through this function. Similarly, there are other functions defined in the smart contract to query the data while describing the requester's requirement.

Through the LUCE platform, the model interacts with Ganache~\cite{ganache}, a test network that creates a virtual Ethereum blockchain and generates pre-configured accounts that are used for the development and testing. The accounts are pre-funded which enables the deployment of the contracts. Ganache provides the balance in \textit{ether} and notifies the \textit{gas} used for running the transactions.

\section{Evaluation}\label{sec:evaluation}
To evaluate our solution we use the D1NAMO dataset~\cite{D1namo}. a publicly available dataset of 29 patients who collected data using wearables and smartphone technology. The dataset contains 20 healthy (H\#001-020) patients and 9 patients with diabetes (D\#001-009). Every participant in the dataset collected data for 3 to 4 days. Among others, these data include heart rate, activity, and breathing collected via a wearable sensor. The 29 participants signed an informed consent form during the collection of the data. The data is now open to everyone, however, the given consent remains unchanged and within the organization that collected the data. Currently, there is no way to meaningfully follow up on the study with the same data subjects. Research clearly suggests that it is natural that the individual consent of the participants evolves as the future needs of researchers studying the data change \cite{kaye2015dynamic, kaye2012patients, chalmers1999people, meslin2010research}. For this reason, we assume that the 29 individuals in the dataset would have different wishes regarding their data use and would provide different consent terms. For example, there might be a diabetic patient who opts to open the data for the research community by expressing a consent statement in the primary category of the DUO model. However, there might be other patients who are not very open to sharing their data with a broad-consent. In this case, they could opt for a \textit{disease-specific} category of consent and share the data only for research on a particular disease.  
\begin{table}
\caption{Data provider profiles for consent representation}
\label{table:dataproviderprofile}
\setlength{\tabcolsep}{3pt}
\def\arraystretch{1.5}
\begin{tabular}{p{130pt}p{100pt}p{100pt}p{100pt}}
\hline
  & Open & Restrictive & Very restrictive \\
\hline
General research & \cmark & \xmark & \xmark \\
\hline
HMB research & \xmark & \cmark & \xmark \\
\hline
DS research & \xmark & \xmark & \cmark \\
\hline
Profit-use & \cmark & \cmark & \xmark \\
\hline
\multicolumn{4}{p{430pt}}{DS= Disease-specific}
\end{tabular}
\end{table}
In order to evaluate our consent model, we consider different data provider profiles and data requesters with different data request statements. 
\subsubsection{Data provider profiles}
Table~\ref{table:dataproviderprofile} represents three data provider profiles. They are classified as \textit{open}, \textit{restrictive} and \textit{very-restrictive} in relation to their consent statement. A data provider consent statement is classified as \textit{open}, if it shares the data for general research use and profit-making purpose. In the \textit{restrictive} data provider category, a data provider refuses to share data for general research, instead, s/he makes the data available for health research and profit-making purpose only. This means that a data requester who requests the data for general research use, is unable to access the data. In the \textit{very-restrictive} category, data providers share the data only for disease-specific research use such as diabetes, cancer, etc. It also opts out from sharing the data for profit-use purposes. Therefore, if a data requester has a profit-making intention, it will be denied access to the data. These are the three major categories where we categorized the data provider consent statements.  
\subsubsection{Data requester profiles}
 Table~\ref{table:datarequesterprofile} represents the data requester profiles divided into three categories. \textit{Data requester 1} is requesting data access for General research use and profit-making intention i.e. it is requesting the data for \textit{method development} or any other research use. \textit{Data requester 2} is requesting data access for HMB research and profit-making intention i.e. it is requesting the data access for \textit{fundamental biology data} or any other data category described in ADA-M matrix form. Similarly, \textit{Data requester 3} is requesting data access for disease-specific research and non-profit use. In the following, we randomly assign a profile to every user. Based on the profile, we define their individual consent and measure what portion of the dataset the data requesters will be able to access given the different individual consent forms in the platform.

\begin{figure}[t!]
\centering
\includegraphics[width=0.9\textwidth]{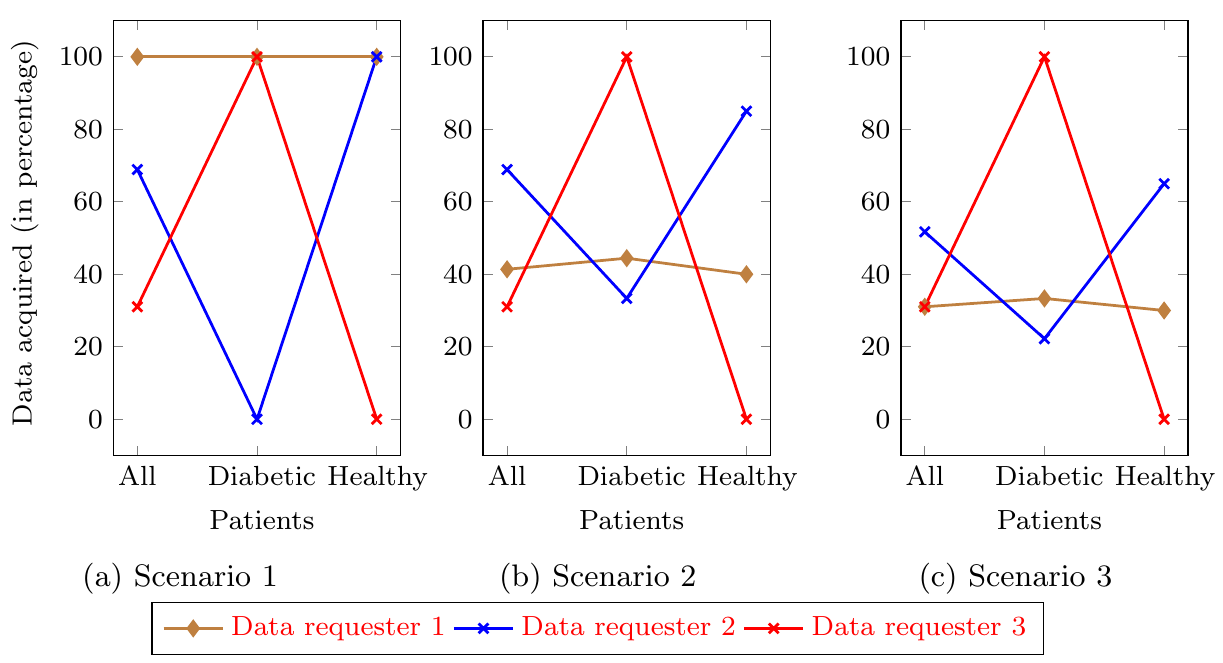}
\caption{Access control evaluation over D1NAMO dataset.}
\label{fig:eval}
\end{figure}

Based on the different data provider profiles defined earlier, We simulate the following scenarios: 
\begin{enumerate}
    \item Scenario 1: $100\%$ open
    \item Scenario 2: $40\%$ open, $40\%$ restrictive, $20\%$ very-restrictive
    \item Scenario 3: $30\%$ open, $30\%$ restrictive, $40\%$ very-restrictive
\end{enumerate}
\begin{table}
\caption{Data requester's profiles for purpose of use statement representation}
\label{table:datarequesterprofile}
\setlength{\tabcolsep}{3pt}
\def\arraystretch{1.5}
\begin{tabular}{p{130pt}p{100pt}p{100pt}p{100pt}}
\hline
  & Data req. 1 & Data req. 2 & Data req. 3 \\
\hline
General research & \cmark & \xmark & \xmark \\
\hline
HMB research & \xmark & \cmark & \xmark \\
\hline
DS research & \xmark & \xmark & \cmark \\
\hline
Profit-use & \cmark & \cmark & \xmark \\
\hline
\multicolumn{4}{p{430pt}}{DS= Disease-specific}
\end{tabular}
\end{table}
Scenario 1 creates consents statements where all the data providers are open to share their data. Thereby all the patients open their data for general research and profit making purpose. Scenario 2 simulates consent statements where $40\%$ of data providers are \textit{open}, while $40\%$ are \textit{restrictive} and rest $20\%$ are \textit{very-restrictive}. Similarly, in the case of scenario 3, data providers submit consent statements where $30\%$ are \textit{open}, while $30\%$ are \textit{restrictive} and rest $40\%$ are \textit{very-restrictive}. This represents the data providers who are more restrictive compared to the previous scenarios but they are open to share more for disease-specific purposes. In this evaluation, we expect that individual consent introduces more individual data sharing control and that the individual dynamics will reflect on which purpose of use statement will be more successful in obtaining the data.\\
\begin{table}
\caption{Consent matching in Scenario 1}
\label{table:consentmatchingScenario1}
\setlength{\tabcolsep}{3pt}
\def\arraystretch{1.5}
\begin{tabular}{p{110pt}p{80pt}p{80pt}p{80pt}p{80pt}}
\hline
   & No. of data provider's & Data req. 1 & Data req. 2 & Data req. 3 \\
\hline
D\#002 & 9 & \cmark & \xmark & \cmark \\
\hline
H\#002 & 20 & \cmark & \cmark & \xmark \\
\hline
\multicolumn{5}{p{430pt}}{DS= Disease-specific}
\end{tabular}
\end{table}
\noindent \textbf{Scenario 1.} In \textit{scenario 1}, everyone is willing to share their data, that means all the healthy and unhealthy subjects are open to share the data with others. Therefore, as seen in Fig.~\ref{fig:eval}, the data requester 1 can access the whole data shared by data providers. Data requester 2 is requesting data for HMB research, therefore it will access only the data from healthy patients, which is the portion of the data needed for this research. Overall, data requester 2 will get the $68.9\%$ of the dataset.  Similarly, the data requester 3 is requesting the data for disease-specific research and non-profit use, therefore, s/he will be able to access the data from diabetic patients as requested by the research purpose statement. This is because all the data providers open their data for profit use and therefore, non-profit data requesters will also get access to the data. We observe a detailed view of consent matching for Scenario 1 in Table~\ref{table:consentmatchingScenario1}.\\
\begin{table}
\caption{Consent matching in Scenario 2}
\label{table:consentmatchingScenario2}
\setlength{\tabcolsep}{3pt}
\def\arraystretch{1.5}
\begin{tabular}{p{100pt}p{80pt}p{80pt}p{80pt}p{80pt}}
\hline
   & No. of data provider's & Data req. 1 & Data req. 2 & Data req. 3 \\
\hline
D\#002 & 4 & \cmark & \xmark & \cmark \\
\hline
H\#002 & 8 & \cmark & \cmark & \xmark \\
\hline
D\#004 & 3 & \xmark & \cmark & \cmark \\
\hline
H\#004 & 9 & \xmark & \cmark & \xmark \\
\hline
D\#005 & 2 & \xmark & \xmark & \cmark \\
\hline
H\#005 & 3 & \xmark & \xmark & \xmark \\
\hline
\multicolumn{5}{p{430pt}}{DS= Disease-specific}
\end{tabular}
\end{table}

\noindent \textbf{Scenario 2.} In \textit{scenario 2}, as seen in Fig.~\ref{fig:eval}, we observe a drop in the percentage of data shared with the data requester 1. It means, based on the consent specified by data providers, there is more access control by the consent model. This is explained by the fact that only $40\%$ of the data providers are open. All the other data providers in this scenario are now restrictive or very-restrictive. We observe that data requester 1 only accesses $41.37\%$ of the dataset, a considerably lower percentage compared to the previous scenario. However, we observe an increase in percentage for data requester 2 that access $68.96\%$ of the dataset. This happens because data requester 2 is requesting data with a more specific purpose of use (HMB research), the $40\%$ of data providers who are restrictive will grant access to more specific research statements (such as HMB research). Similarly, for data requester 3, we don't see any changes in the dataset access percentage. Since data requester 3 is seeking the data for disease-specific and non-profit use, therefore, it will always match the consent, even from the restrictive user profiles. We observe a detailed view of consent matching results for Scenario 2 in Table~\ref{table:consentmatchingScenario2}.
\\
\begin{table}
\caption{Consent matching in Scenario 3}
\label{table:consentmatchingScenario3}
\setlength{\tabcolsep}{3pt}
\def\arraystretch{1.5}
\begin{tabular}{p{100pt}p{80pt}p{80pt}p{80pt}p{80pt}}
\hline
   & No. of data provider's & Data req. 1 & Data req. 2 & Data req. 3 \\
\hline
D\#002 & 3 & \cmark & \xmark & \cmark \\
\hline
H\#002 & 6 & \cmark & \cmark & \xmark \\
\hline
D\#004 & 2 & \xmark & \cmark & \cmark \\
\hline
H\#004 & 7 & \xmark & \cmark & \xmark \\
\hline
D\#005 & 4 & \xmark & \xmark & \cmark \\
\hline
H\#005 & 7 & \xmark & \xmark & \xmark \\
\hline
\multicolumn{5}{p{430pt}}{DS= Disease-specific}
\end{tabular}
\end{table}
\noindent \textbf{Scenario 3.} In \textit{scenario 3}, as seen in Fig.~\ref{fig:eval}, we observe a sharing percentage drop for data requester 1 and 2 compared to the previous scenarios. Since only $30\%$ data providers are open while the majority of data providers are very-restrictive, there is a strict access control by the consent model. We observe that data requester 1 can only access $31.03\%$ of the dataset while data requester 2 can only access $51.72\%$ of the dataset. We observe no changes in percentage for data requester 3 which is explained by the profile of data requester 3 which is requesting data for disease-specific research.  We observe a detailed view of consent matching results for Scenario 3 in Table~\ref{table:consentmatchingScenario3}.

Fig.~\ref{fig:eval}, shows the evaluation results in the three different scenarios. The results show that our consent model for data sharing gives users greater data control and that the different user consents will reflect on how specific the data requesters are in their purpose of use statements. 

\subsubsection{Cost Analysis} 
\begin{figure}[t!]
\centering
\includegraphics[width=0.9\textwidth]{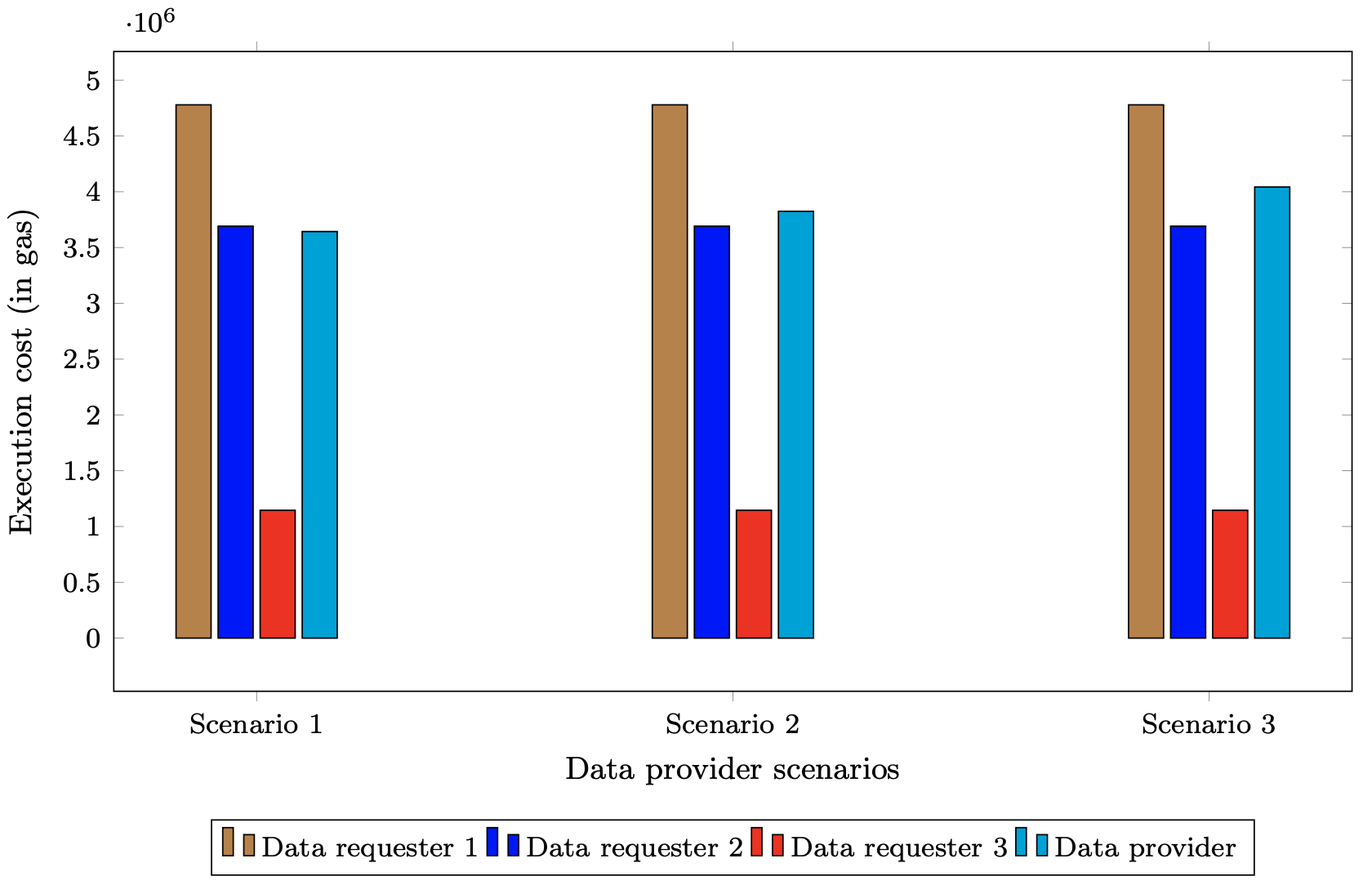}
\caption{Gas consumption in different scenarios for whole dataset.}
\label{fig:gas1}
\end{figure}

To understand the applicability of the consent model, we perform a cost analysis to evaluate the applicability of our smart contract model. In performing the cost analysis, we considered several parameters:
\begin{enumerate}
\item Total amount of gas spent during contract deployment.
\item Amount of gas consumed on executing the data provider consent.
\item Amount of gas consumed on executing the data requester queries.
\item Amount of gas spent while matching the consent.
\end{enumerate}
In the Ethereum environment, every operation in the smart contract consumes some amount of gas (a measure of the computational effort required to perform an operation). Some operations require more computational effort than others, therefore the gas consumption can vary. We evaluate gas consumption of the smart contract based on the relevant operations and based on whom is performing them (the data provider or the data requester). Table~\ref{table:costanalysis} shows the total gas consumption to execute the contract is 1765926. We considered the average gas price of 8 Gwei according to the current date and time of ETH gas station~\cite{gasstation}. Therefore, the relevant cost of contract deployment is 0.0141274 ETH with a corresponding price of \$ 2.92. \\
To better understand the involved costs, we calculate the total cost of contract deployment and the cost of interactions between the data providers and data requesters. Therefore, the total cost for sharing the data is measured by the accumulative cost of contract deployment, the cost for submitting the consent, and the data requesters' cost for querying the data. Fig.~\ref{fig:gas1} shows the amount of gas consumption for different scenarios. From the results, we can observe that most data requesters consume more gas than data providers. This is explained by the number of interactions that they have with the smart contracts of data providers. The broader the expressed purpose of the use is, the more data providers may be contacted to ask for data access. This is also reflected by the fact that the data requester 3, who is looking only for diabetes data, will be requesting the data only to 9 of the 29 participants as opposed to data provider 1 and 2 who will contact all the 29 participants. 
Furthermore, data providers who are more restrictive, would have to evoke additional functions in the smart contract than the more open data requesters, this reflects into a slight increase in costs for data providers by approximately by 6 percent amounting to \$ 0.006. A detailed table of the costs is shown in ~\ref{table:costanalysis}. It is possible to fund the data provider costs by requesting a small fee to be paid by the data requesters when the data is accessed. The overall costs in the system will remain the same, however, the incentives between data requesters and data providers could be better balanced if data requesters do not occur costs to share their data.
The deployment cost for one smart contract is $\approx$ \$ 3 and this is covered once by the data provider. The total deployment cost of the whole D1NAMO dataset amounts to \$87. To place this into perspective, the costs of the D1NAMO dataset project approximately several thousands more. While these studies are important, the data collection could be achieved by our model with minor efforts and with more flexibility for secondary uses of the same data.\\

\begin{figure}[t!]
\centering
\includegraphics[width=0.9\textwidth]{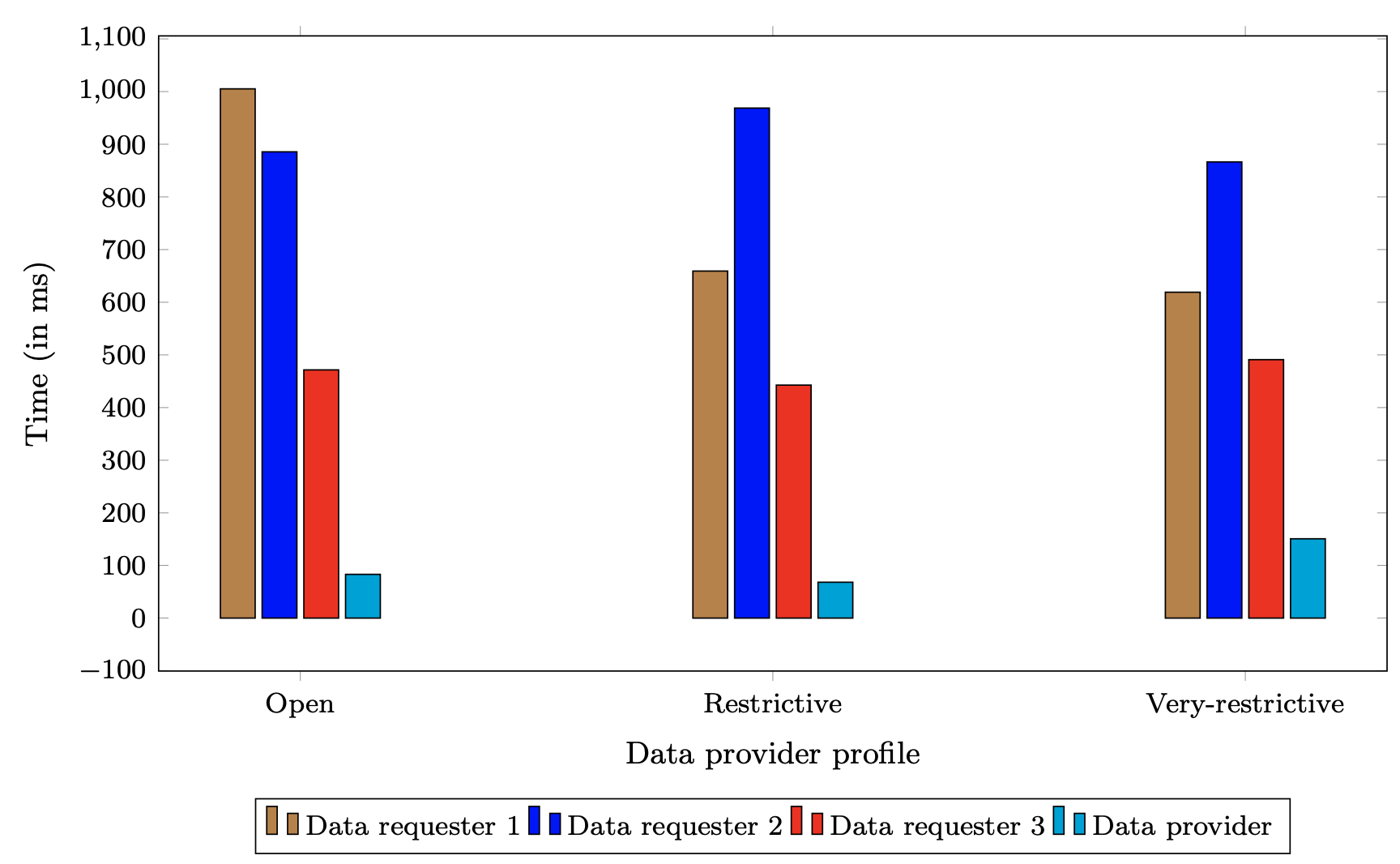}
\caption{Latencies of data providers and data requesters.}
\label{fig:latency1}
\end{figure}

\noindent\textbf{Latency measurements.} With an increasing willingness to share data, we believe that individual data would be collected in much shorter time frames. While platforms like LUCE are not available to the public yet, we show that it would be possible to efficiently reconstruct databases even in scenarios where users would not be always fully open to data sharing. For example, if we want to develop the same dataset of 29 patients, each patient can easily share the data on the LUCE platform by submitting their consent. As seen in Fig.~\ref{fig:latency1}, the latency of a data requester is higher due to the interaction with multiple contracts. The single interaction with one smart contract is in the order of 50 ms. This means that for interacting with 1000 potential data providers, it would take about 50 seconds. The interaction with all the 29 patients in the dataset, it takes 1.5 seconds. Fig.~\ref{fig:latency1} shows the average interaction by each data requester which is $\approx$ 1 sec. Moreover, it takes an average of 220 ms to submit queries by each data requester.
The latency for data providers might vary. For example, a very-restrictive data provider uses the  \textit{UploadDataRequirements} function, which takes $\approx$ 150.6 ms to execute.

\begin{table}
\caption{Gas consumption of smart contract}
\label{table:costanalysis}
\setlength{\tabcolsep}{3pt}
\def\arraystretch{1.5}
\begin{tabular}{p{110pt}p{80pt}p{80pt}p{80pt}p{80pt}}
\hline
 Actions & Transaction cost & Execution cost &  Ether cost & Cost* \\
\hline
Deployment & 2372326 & 1765926 & 0.0141274 & \$ 2.92437 \\
\hline
UploadData\\PrimaryCategory & 88129 & 64745 & 0.000518 & \$ 0.10567 \\
\hline
UploadData\\SecondaryCategory & 26814 & 3622 & 0.000029 & \$ 0.00592 \\
\hline
UploadData\\Requirements & 60053 & 36285 & 0.0002903 & \$ 0.05922 \\
\hline
giveResearch\\Purpose & 87280 & 63896 & 0.0005112 & \$ 0.10428\\
\hline
giveHMBPurpose & 45902 & 22198 & 0.0001776 & \$ 0.03623\\
\hline
giveClinical\\Purpose & 44699  & 21635 & 0.0001731 & \$ 0.03531\\
\hline
giveGeographic\\SpecificRestriction & 25182 & 2374 & 0.000019 & \$ 0.00388\\
\hline
giveProfit & 44723 & 21723 & 0.0001738 & \$ 0.03546 \\
\hline
givePerson & 45357 & 22101 & 0.0001768 & \$ 0.03606 \\
\hline
giveData\\RequesterTerms & 46265 & 22177 & 0.0001774 & \$ 0.03619 \\
\hline
AccessData (Max) & 84341 & 60253 & 0.000482 & \$ 0.09833  \\
\hline
AccessData (Min) & 32543 & 8455 & 0.0000676 & \$ 0.01379  \\
\hline
\multicolumn{5}{p{430pt}}{*= Ether conversion with present date price (Average (8 Gwei))}
\end{tabular}
\end{table}
\section{Discussion}\label{sec:discussion}
In this section, we discuss several issues related to data sharing using a blockchain-based consent model.
\subsection{Scalability} Our blockchain-based consent model gives more flexibility than current methods where organizations must ask for consent using physical forms. Also, with the help of a smart contract, there is no need for a trusted third party to process the data, hence, it reduces the cost of sharing the data between the organizations or individuals. Using this model, we can have a large data collection with fast processing and in limited time. Currently, there is a lot of data that is unused because of unfair data sharing practices. With our model, We expect more individual adherence to data sharing. Scalability in blockchain platforms can become a challenge when the number of users increases~\cite{vukolic2016scalable} and in our future works, we will seek to measure how well the platform scales with a large number of users. 
\subsection{Data heterogeneity for data sharing} 
Shared data are heterogeneous in nature. When sharing data in our platform, there is a need for common ontological representation models where data providers can describe their contents related to their consent. Solutions such as OHDSI~\cite{ohdsi} and other health data ontological solutions can help to better represent the contents of the shared data.
\subsection{Data privacy and access control}
In our consent model, individuals are enabled to express consent over the use of their data use and change, modify or request to delete all data if desired. This ensures compliance with privacy protection laws such as the GDPR. Moreover, the transaction data stored in the blockchain do not reveal patient identities nor contain the records of patients. They are rather transactions containing the information about who accessed what records for which purpose. However, there is a need for a detailed studies on how these models can be enrolled in a secure and privacy preserving manner~\cite{sieve,kanonymity2002,robustanonymization2008,bpsFGCS}. The security and privacy aspects were not the focus of this work, we believe however that they are imperative for the acceptability of such solutions among users. In our future works, we will further investigate the use of cryptographic protocols, such as multi-party computation \cite{yao1986generate}, and token-generation \cite{sharma2017software} for further securing the data-sharing platform.
\subsection{Authorization}
In the LUCE platform, data providers and data requesters register themselves before sharing data. Currently, authorization but not strong authentication is provided in LUCE. Yet, the verification of identity is possible to avoid unwanted users on the platform. LUCE can be easily extended to verify identity (such as a passport or any other national ID card) of data providers and requesters at the moment of registration. A second point of attack in public blockchain networks is if a malicious user bypasses the registration, identifies the smart contract address and requests access to the data without being a registered LUCE platform user.  For this purpose, we link the token mechanisms for assessing the end-point data to the authentication service. This means that only authorized users will be given access to the data. In our work, we are currently investigating the use of OAuth 2.0~\cite{oAuth}~\cite{oauthVasilios} protocol with blockchain tokens to define secured tokens from data providers. These secure tokens enable data requesters to access a protected resource with the consent of the resource owner. This access is given by an authorization server in terms of \textit{access tokens}. The details of this work will be described in detail in our future publications.
\section{Conclusion \& Future Work}\label{sec:conclusion}
Data privacy is an important issue for most of the data providers. In the data sharing context, data requesters must comply with GDPR and the consent of data providers. We found out that work on facilitating data sharing with user consent is in an early stage. In this paper, we proposed a framework for sharing data with explicit consent form data providers. We use the DUO and ADA-M model to embed a standardized ontological consent and data request model. We develop a smart contract to automate a generic consent model, we deploy and test the solution in LUCE, our data sharing platform. The evaluation shows that data providers have more data control while sharing the data on a blockchain.

As future work, we plan to investigate the scalability of our model where we would like to see the impact in real-life scenarios where the number of data providers or data requesters is very large. This could impact the performance of sharing which is a current working issue in any blockchain platform. Also, we would like to simulate large tests to ensure that the consent mechanism maintains privacy and be acceptable among data providers for data sharing. 

\section*{Acknowledgment}
We would like to thank Anthony Krieger for his initial insight on the work which helped us to complete this work on time. We also thank Arno Angerer for his initial implementation of the LUCE platform which helped us to develop our model.

\nocite{*} 

\end{document}